\documentclass[conference]{IEEEtran}
\IEEEoverridecommandlockouts

\usepackage{cite}
\usepackage{amsmath,amssymb,amsfonts}
\usepackage{algorithmic}
\usepackage{graphicx}
\usepackage{textcomp}
\PassOptionsToPackage{table}{xcolor}
\usepackage{xcolor}

\usepackage{booktabs}
\usepackage{multirow}
\usepackage{graphicx}
\usepackage{subcaption} 

\definecolor{TableGray}{gray}{0.9}

\DeclareMathOperator*{\argmin}{arg\,min}

\newcommand{\bc}[1]{\mbox{\boldmath $\mathcal{#1}$}}
\newcommand{\mf}[1]{\mathbf{#1}}
\newcommand{\mb}[1]{\mathbb{#1}}
\newcommand{\bs}[1]{\boldsymbol{#1}}

\def\BibTeX{{\rm B\kern-.05em{\sc i\kern-.025em b}\kern-.08em
		T\kern-.1667em\lower.7ex\hbox{E}\kern-.125emX}}
\usepackage{hyperref}
\begin{document}

	\title{Deep Tensor Learning for Reliable Channel Charting from Incomplete and Noisy Measurements
	}

	\author{\IEEEauthorblockN{Ge Chen, Panqi Chen and Lei Cheng}
		\IEEEauthorblockA{\textit{College of Information Science and Electronic Engineering} \\
			\textit{Zhejiang University, Hangzhou, 310027, China} \\
			{gechen@zju.edu.cn, 12431109@zju.edu.cn, lei\_cheng@zju.edu.cn}}
			\thanks{Corresponding author: Lei Cheng (lei\_cheng@zju.edu.cn). This work was supported by the National Natural Science Foundation of China under Grant 62371418.}
	}
	
	\maketitle

	\begin{abstract}
		Channel charting has emerged as a powerful tool for user equipment localization and wireless environment sensing. Its efficacy lies in mapping high-dimensional channel data into low-dimensional features that preserve the relative similarities of the original data. However, existing channel charting methods are largely developed using simulated or indoor measurements, often assuming clean and complete channel data across all frequency bands. In contrast, real-world channels collected from base stations are typically incomplete due to frequency hopping and are significantly noisy, particularly at cell edges. These challenging conditions greatly degrade the performance of current methods. To address this, we propose a deep tensor learning method that leverages the inherent tensor structure of wireless channels to effectively extract informative while low-dimensional features (i.e., channel charts) from noisy and incomplete measurements. Experimental results demonstrate the reliability and effectiveness of the proposed approach in these challenging scenarios. 
	\end{abstract}
	
	
	\section{Introduction}
	Modern wireless systems are expected to support denser user equipments (UEs), larger antenna arrays, and an increased number of subcarriers\cite{zeng2024tutorial}. These trends lead to growing volumes of high-dimensional channel data collected at base stations (BSs), creating new opportunities for intelligent data-driven learning methods to extract knowledge from these large datasets and redefine the potential of wireless sensing, localization, and communication. 
	
	To this end, an emerging technique called {\it channel charting}\cite{studer2018channel,ferrand2021triplet,ferrand2023wireless,stephan2024angle}maps high-dimensional channel data into low-dimensional representations (also known as pseudo-positions) that preserve the relative similarities of the original data. With a properly designed similarity measure, the resulting low-dimensional channel chart can provide valuable insights into the radio environment, such as user positions/trajectories within a region, and the clustering properties of channels, facilitating various downstream tasks\cite{le2022channel,kazemi2021channel}.
	
	Earlier works in channel charting primarily rely on classical manifold learning algorithms, such as Multidimensional Scaling (MDS)\cite{kruskal1964multidimensional} and t-Distributed Stochastic Neighbor Embedding (t-SNE)\cite{van2008visualizing}. However, these methods are mostly non-parametric, making out-of-sample extensions computationally expensive. Specifically, obtaining the low-dimensional representations (i.e., pseudo-positions) for new channel data requires referencing historical data\cite{bengio2003out}, which is inefficient. To address this limitation, recent approaches leverage deep learning-based\cite{ferrand2021triplet,lei2019siamese} dimensionality reduction techniques to learn a parametric mapping, enabling efficient out-of-sample predictions. Among these, the most representative methods are based on Siamese or Triplet network architectures\cite{ferrand2021triplet,lei2019siamese}. Their key idea is to train a neural network that takes channel data as input and outputs pseudo-positions, ensuring that the relative dissimilarities among these pseudo-positions closely match those computed from the high-dimensional channel data. Once trained, the neural network can rapidly generate low-dimensional pseudo-positions for any new channel samples, enabling fast channel charting.    
	
	Despite their effectiveness, current deep learning-based channel charting methods often {\it overlook non-ideal} factors encountered in practical scenarios, such as noisy communication environments and incomplete channel measurements (e.g., due to frequency hopping)\cite{liu2011frequency,wu2021frequency}. Additionally, existing feature extraction techniques for channel data tend to retain excessive information, leading to high-dimensional feature vectors (e.g., computed via auto-correlation)\cite{stephan2024angle}. As a result, neural networks predominantly rely on Multi-Layer Perceptrons (MLPs), significantly increasing the number of parameters and the training burden. These challenges raise a critical question: {\it How can we design a neural network-based channel charting method that is robust to noisy and incomplete channel data while maintaining a concise model architecture to enable fast training and inference?}
	
	To address this question,  we make {\it the first attempt} to leverage the inherent {\it low-rank tensor structure} within wireless channel data to design a robust and efficient channel charting network. Recent studies have shown that wireless channels in Multiple-Input Multiple-Output Orthogonal Frequency Division Multiplexing (MIMO-OFDM) systems can be naturally represented using low-rank tensor decomposition models\cite{araujo2019tensor,zhou2017low,zhang2022tensor,zhang2024channel}. By leveraging tensor structures, more accurate channel estimations have been achieved while requiring fewer pilot signals. These findings motivate us to develop a charting network that integrates {\it tensor decompositions and computations} to mitigate the adverse effects of incomplete and noisy channel measurements, while reducing the number of network parameters. Experimental results demonstrate the superior performance of our proposed method compared to state-of-the-art approaches.
	
\flushbottom

	\section{Problem Formulation And Challenges Ahead}
	\subsection{Channel Model}
	We consider a MIMO-OFDM system. Following\cite{3gpp}, the channel between the BS and the UE is modeled using the 3GPP Cluster Delay Line (CDL) model:
	\begin{align}\label{eq:CDL_model}
		h(i_{\text{pol}}, i_{\text{tx}}, i_{\text{rx}}, i_{\text{sc}}) = \sum_{l=0}^{L-1} \alpha_{l, i_{\text{pol}}} e^{-j2\pi \tau_l(i_{\text{sc}} \Delta f)} \notag \\
		\times e^{j2\pi f_c  \frac{\hat{\bf r}_{\text{tx}, l}^T \bar{\bf d}_{i_{\text{tx}}}}{c}}
		e^{j2\pi f_c  \frac{\hat{\bf r}_{\text{rx}, l}^T \bar{\bf d}_{i_{\text{rx}}}}{c}}.
	\end{align}
	Here, $h(i_{\text{pol}}, i_{\text{tx}}, i_{\text{rx}}, i_{\text{sc}})$ denotes the channel coefficient, where $i_{\text{pol}}, i_{\text{tx}}, i_{\text{rx}}$ and $i_{\text{sc}}$ represent the polarization index, UE transmit antenna index, BS receive antenna index, and subcarrier index, respectively. The channel coefficient $h(i_{\text{pol}}, i_{\text{tx}}, i_{\text{rx}}, i_{\text{sc}})$  can be interpreted as the sum of the responses from  $L$  sub-paths.

	In the $l$-th sub-path, $\alpha_{l, i_{\text{pol}}}$ is the complex gain corresponding to the $i_{\text{pol}}$-th polarization, and $\tau_l$ is the propagation delay. The constants $\Delta f$, $f_c$, and $c$ 
	denote the subcarrier spacing, carrier frequency, and speed of light, respectively.
	
	The spherical unit vector $\hat{\bf r}_{\text{rx}, l}$, defined by the azimuth angle of arrival $\phi_{l, \text{AOA}}$ and zenith angle of arrival $\theta_{l, \text{ZOA}}$, is:
	\begin{equation}
		\hat{\bf r}_{\text{rx}, l}  \stackrel{\Delta}{=}
		\begin{bmatrix}
			\text{sin} \theta_{l, \text{ZOA}}\text{cos} \phi_{l, \text{AOA}} \\
			\text{sin} \theta_{l, \text{ZOA}}\text{sin} \phi_{l, \text{AOA}} \\
			\text{cos} \theta_{l, \text{ZOA}}
		\end{bmatrix}.
	\end{equation}
	Similarly, $\hat{\bf r}_{\text{tx}, l}$ is determined by the azimuth angle of departure  $\phi_{l, \text{AOD}}$ and  the zenith angle of departure $\theta_{l, \text{ZOD}}$. The vectors $\bar{\bf d}_{i_{\text{tx}}}$ and $\bar{\bf d}_{i_{\text{rx}}}$ specify the three-dimensional  Cartesian coordinates of the $i_{\text{tx}}$-th UE antenna and the $i_{\text{rx}}$-th BS antenna, respectively.

	Note that $h(i_{\text{pol}}, i_{\text{tx}}, i_{\text{rx}}, i_{\text{sc}})$ depends on multiple indices; therefore, it can be naturally represented as a channel tensor $\bc{H} \in \mathbb{C}^{N_{\text{rx}}\times N_{\text{pol}}\times N_{\text{tx}}\times N_{\text{sub}}}$, where $N_{\text{rx}}, N_{\text{pol}}, N_{\text{tx}}$, and $N_{\text{sub}}$ denote the number of BS antennas, polarization modes, UE antennas, and subcarriers, respectively.

	\subsection{Channel Charting and Challenges Ahead}\label{AA}
	Channel charting can be expressed as a mapping function:
	\begin{align}\label{formula(3)}
		\mathbf z (\bc {H}_i):   \mathbb{C}^{N_{\text{rx}} \times N_{\text{pol}} \times N_{\text{tx}}\times  N_{\text{sub}}} \rightarrow \mathbb{R}^D
	\end{align}
	which projects the high-dimensional channel $\bc{H}_i$ into a lower-dimensional pseudo-position $\mathbf{z} (\bc{H}_i)$, where $D \ll N_{\text{rx}} N_{\text{pol}} N_{\text{tx}} N_{\text{sub}}$.
	
	The goal is to preserve the relative similarity between channels, meaning that the distance between $\mathbf{z} (\bc{H}_i)$ and $\mathbf{z} (\bc{H}_j)$ should reflect the dissimilarity $d_{ij} \triangleq d(\bc{H}_i, \bc{H}_j)$ in the original space. Accordingly, channel charting can be formulated as the optimization problem:
	\begin{align}\label{optimization problem}
		\min_{\mathbf z(\cdot)} \sum_{i=1}^{N-1}\sum_{j=i+1}^{N} 
		\Bigl(d_{ij} - \|\mathbf  z (\bc {H}_i) - \mathbf  z (\bc {H}_j) \|_2\Bigr)^2. 
	\end{align}
	
	To address this problem, various charting algorithms have been proposed, broadly categorized into manifold learning-based and deep learning-based approaches\cite{ferrand2023wireless,stephan2024angle}. We focus on the latter due to their ability to learn a parametric mapping $\mf{z}_{\bs \theta} (\cdot)$, which enables efficient out-of-sample predictions, where $\bs \theta$ represents the neural network parameters.
	
	A major challenge in deep learning-based methods is the lack of direct support for complex-valued inputs in standard frameworks (e.g., PyTorch). To address this, carefully designed channel features are often used to encode channel properties for training. Most existing feature representations rely on auto-correlation\cite{stephan2024angle} in the time or angular domain\cite{studer2018channel}, resulting in extremely high-dimensional features. Consequently, neural networks processing these flattened features—typically implemented as MLPs—incur a high parameter count and significant training overhead. Furthermore, current methods seldom account for real-world impairments such as noisy or incomplete measurements. {\it This limitation reduces their robustness and efficiency  in practical deployments, highlighting the need for more resilient feature representations and concise  neural architectures}.
	
\section{Proposed Approach}\label{Section III}
\begin{figure*}[!t]
	\vspace{-2em}
	\centering
	\includegraphics[width=\textwidth]{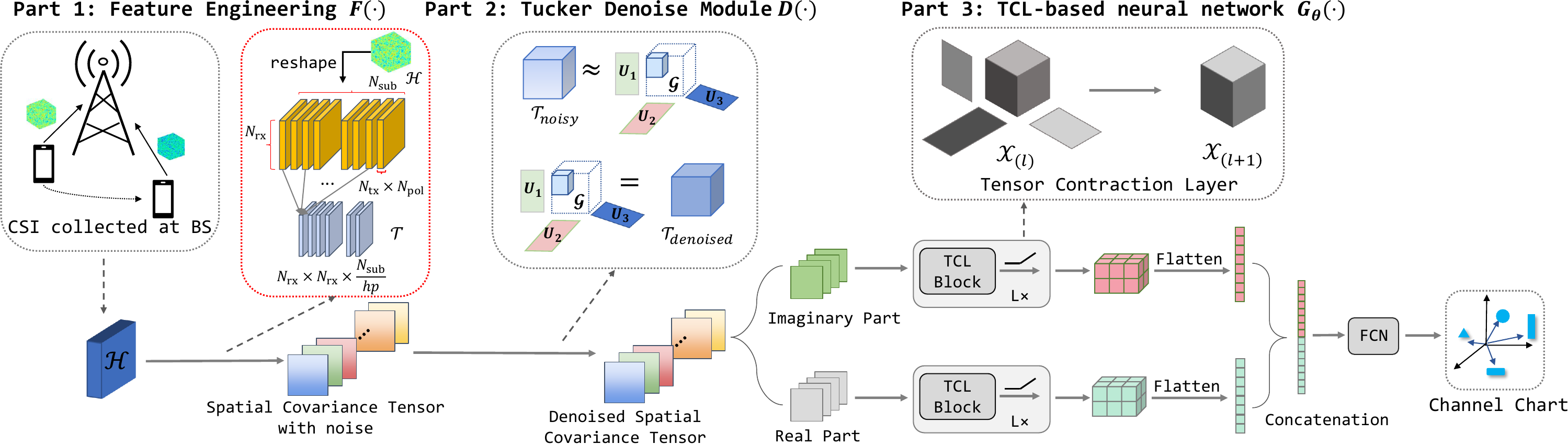}
	\caption{The detailed architecture of our proposed deep tensor learning model for reliable channel charting.}
	\label{fig:tdr_model}
	\vspace*{-1em}
\end{figure*}

In this section, we introduce our \emph{deep tensor learning} (DTL) framework for reliable channel charting. Similar to existing methods, DTL begins with a feature engineering module $F(\cdot)$  that transforms raw channel data  $\bc H$ into a more robust tensor representation $F(\bc H)$, mitigating real-world impairments. A tensor Tucker decomposition-based denoising module $D(\cdot)$ then further suppress noise\cite{kolda2009tensor}. Unlike prior approaches relying on MLP-based dimensionality reduction, our framework employs a \emph{tensor contraction layer}\cite{kossaifi2017tensor,kossaifi2020tensor,li2023striking} operating directly on tensor-based features to produce the final channel chart. Fig.~\ref{fig:tdr_model} illustrates the overall architecture.

\subsection{Part 1: Tensor Feature Engineering }
The feature engineering module  $F(\cdot)$  is crucial yet challenging in channel charting. It must strike a balance between expressiveness and conciseness while ensuring robustness. Retaining too much raw data increases feature dimensions and network parameters, while overly compact features risk losing essential channel details, degrading chart quality.

Towards these goals, we propose computing a \emph{spatial covariance tensor} \(\boldsymbol{\mathcal{T}}  \in \mathbb{C}^{N_{\text{rx}} \times N_{\text{rx}} \times \tfrac{N_{\text{sub}}}{h_p}}\) for each channel tensor \(\boldsymbol{\mathcal{H}}\), as illustrated in Part 1 of Fig.~\ref{fig:tdr_model}. Concretely, to mitigate the impact of incomplete channel information along the subcarrier dimension due to frequency hopping, we assume a $\tfrac{1}{h_p}$ frequency hopping, meaning each observation captures only  \(\tfrac{1}{h_p}\)  of the total subcarriers. As illustrated in the red box of Fig.~\ref{fig:tdr_model}, we select every $h_p$ channel slice matrices to form \(\tfrac{N_{\text{sub}}}{h_p}\) subtensors  along the subcarrier mode:
\begin{equation}\label{formula(7)}
	\boldsymbol{\mathcal{H}}_i = \boldsymbol{\mathcal{H}}(:, :, :, \mathcal{J}_i) \in \mathbb{C}^{N_{\text{rx}}\times N_{\text{pol}}\times N_{\text{tx}}\times h_p}, i=1, ...,\tfrac{N_{\text{sub}}}{h_p},
\end{equation}
where the index set of the \(i\)-th group is:
\begin{equation}\label{formula(6)}
	\mathcal{J}_i = \Bigl\{ i+\tfrac{N_{\text{sub}}}{h_p}(k-1) \mid k=1, \dots, h_p \Bigr\}, i=1,\dots, \tfrac{N_{\text{sub}}}{h_p}.
\end{equation}

For the $i$-th subtensor, we perform a \emph{mode-1 unfolding} of \(\boldsymbol{\mathcal{H}}_i\) to obtain 
\(\mathbf{H}_{(i),1} \in \mathbb{C}^{N_{\text{rx}} \times (N_{\text{pol}}\,N_{\text{tx}}\,h_p)}\). We then compute its covariance matrix along the first mode (known as the spatial covariance matrix):
\begin{align}\label{formula(7)}
\mf C_i = \tfrac{1}{N_{\text{pol}}\,N_{\text{tx}}\,h_p}
	\,\mathbf{H}_{(i),1}\,\mathbf{H}_{(i),1}^H
	\;\in\; 
	\mathbb{C}^{N_{\text{rx}} \times N_{\text{rx}} }.
\end{align}
Finally, we stack these spatial covariance matrices to construct the spatial covariance  \(\boldsymbol{\mathcal{T}}
\in \mb C^{N_{\text{rx}} \times N_{\text{rx}} \times \tfrac{N_{\text{sub}}}{h_p} } \), i.e., $\boldsymbol{\mathcal{T}}(:,:,i)
 = \mf C_i$.

The proposed spatial covariance tensor $\boldsymbol{\mathcal{T}}$ captures informative spatial correlations among BS antennas across different subbands, {\it making it expressive}. Its dimensionality is significantly smaller than that of the original channel tensor or conventional auto-correlation feature vectors, {\it ensuring conciseness}. Furthermore, partial observations (due to frequency hopping) still allow the computation of $\boldsymbol{\mathcal{T}}$, and the averaging operation in~\eqref{formula(7)} helps mitigate noise to some extent. Therefore, $\boldsymbol{\mathcal{T}}$ is also {\it robust} against noise and missing channel data.

\subsection{Part 2: Tucker Denoise Module}
We begin by recalling tensor Tucker decomposition\cite{kolda2009tensor}. For a third-order tensor 
\(\boldsymbol{\mathcal{Y}} \in \mathbb{C}^{I_1 \times I_2 \times I_3}\), 
its Tucker representation is
\begin{equation}\label{formula(9)}
	\boldsymbol{\mathcal{Y}} 
	\approx 
	\boldsymbol{\mathcal{G}}
	\times_1 \mathbf{U}_1
	\times_2 \mathbf{U}_2
	\times_3 \mathbf{U}_3,
\end{equation}
where \(\mathbf{U}_n \in \mathbb{C}^{I_n \times R_n}\) (\(n = 1,2,3\)) are orthonormal factor matrices (\(\mathbf{U}_n^H \mathbf{U}_n = \mathbf{I}_n\)), \(R_n \le I_n\), and 
\(\boldsymbol{\mathcal{G}} \in \mathbb{C}^{R_1 \times R_2 \times R_3}\) is the core tensor. Here, \(\times_l\) denotes a mode-\(l\) product: for an \(L\)-th order tensor 
\(\boldsymbol{\mathcal{A}} \in \mathbb{C}^{I_1 \times \dots \times I_L}\) 
and a matrix 
\(\mathbf{B} \in \mathbb{C}^{J_l \times I_l}\), 
the result 
\(\boldsymbol{\mathcal{C}} = \boldsymbol{\mathcal{A}} \times_l \mathbf{B}\) 
satisfies
\begin{equation}
	\boldsymbol{\mathcal{C}}_{i_1,\dots,j_l,\dots,i_L}
	\;=\; 
	\sum_{k=1}^{I_l}
	\boldsymbol{\mathcal{A}}_{i_1,\dots,k,\dots,i_L}
	\,\mathbf{B}_{j_l,k}.
\end{equation}

Similar to SVD, choosing relatively small \(\{R_n\}_{n=1}^3\) in Tucker decomposition helps filter out noise:
\begin{equation}\label{formula(10)}
	\boldsymbol{\mathcal{Y}}_{\text{recons}}
	\;=\;
	\boldsymbol{\mathcal{G}}
	\times_1 \mathbf{U}_1
	\times_2 \mathbf{U}_2
	\times_3 \mathbf{U}_3.
\end{equation}
Motivated by this principle, we apply~\eqref{formula(9)} and~\eqref{formula(10)} to 
\(\boldsymbol{\mathcal{T}}\)
to obtain 
\(\boldsymbol{\mathcal{T}}_{\text{denoised}}\), as illustrated in Part 2 of Fig.~\ref{fig:tdr_model}. 
For neural network training, we then separate its real and imaginary parts, 
\(\Re(\boldsymbol{\mathcal{T}}_{\text{denoised}})\) 
and 
\(\Im(\boldsymbol{\mathcal{T}}_{\text{denoised}})\), 
which serve as inputs to subsequent layers. We refer to this entire process as the \emph{Tucker denoise module}, denoted by \(D(\cdot)\).

\subsection{Part 3: Integrating Tensor Computations in the Neural Network}\label{SCM}
Our framework employs a \emph{tensor contraction layer} (TCL)\cite{kossaifi2017tensor,kossaifi2020tensor} to process the feature tensors, as illustrated in Part 3 of Fig.~\ref{fig:tdr_model}. Let 
\(\boldsymbol{\mathcal{X}}_{(l)} \in \mathbb{R}^{I_1^l \times I_2^l \times I_3^l}\) 
be the input to the \(l\)-th TCL. Its output is
\begin{equation}
	\boldsymbol{\mathcal{X}}_{(l+1)} 
	\;=\;
	\zeta\!\Bigl(
	\boldsymbol{\mathcal{X}}_{(l)} 
	\times_1 \mathbf{V}_{(l)}^{(1)}
	\times_2 \mathbf{V}_{(l)}^{(2)}
	\times_3 \mathbf{V}_{(l)}^{(3)}
	\Bigr),
\end{equation}
where 
$\{ \mathbf{V}_{(l)}^{(i)}\in \mathbb{R}^{I_i^{l+1}\times I_i^l} \}_{i=1}^{3} $
are learnable factor matrices, 
$\boldsymbol{\mathcal{X}}_{(l+1)} \in \mathbb{R}^{I_1^{(l+1)} \times I_2^{(l+1)} \times I_3^{(l+1)}}$,
and \(\zeta(\cdot)\) is activation function.

Because TCL builds upon Tucker decomposition, it effectively captures multi-dimensional structures with relatively few parameters. To further increase model capacity, we stack multiple TCLs to form an \(L\)-layer block:
\begin{align}
	\boldsymbol{\mathcal{X}}_{(L)} 
	&= 
	\zeta\Bigl(\dots 
	\zeta\bigl(
	\boldsymbol{\mathcal{X}}_{(0)}
	\times_1 \mathbf{V}_{(0)}^{(1)}
	\times_2 \mathbf{V}_{(0)}^{(2)}
	\times_3 \mathbf{V}_{(0)}^{(3)}
	\bigr)
	\dots \notag\\
	&\quad
	\times_1 \mathbf{V}_{(L-1)}^{(1)}
	\times_2 \mathbf{V}_{(L-1)}^{(2)}
	\times_3 \mathbf{V}_{(L-1)}^{(3)}
	\Bigr),
\end{align}
where 
\(\boldsymbol{\mathcal{X}}_{(0)}\) 
is the initial tensor input and 
\(\{\mathbf{V}_{(i)}^{(j)} | i=0,\dots,L-1;\, j=1,2,3\}\) 
are learnable factors. Let 
\(\boldsymbol{\mathcal{X}}_{\text{Re},(L)}\) 
and 
\(\boldsymbol{\mathcal{X}}_{\text{Im},(L)}\) 
denote outputs of these two \(L\)-layer TCL blocks with inputs \(\Re(\boldsymbol{\mathcal{T}}_{\text{denoised}})\) 
and 
\(\Im(\boldsymbol{\mathcal{T}}_{\text{denoised}})\). We vectorize and concatenate them as
\begin{equation}
	\text{Concat}(\text{Vec}(\boldsymbol{\mathcal{X}}_{\text{Re},(L)}), \text{Vec}(\boldsymbol{\mathcal{X}}_{\text{Im, (L)}})) \in \mathbb{R}^{2(I_1^L\times I_2^L \times I_3^L)},
\end{equation}
which is finally fed into a fully connected network (FCN) to produce the channel chart. We denote the overall parametric mapping as \(G_{\boldsymbol{\theta}}(\cdot)\), where $\boldsymbol{\theta}$ denotes the set of parameters in TCL layers and FCN.

\subsection{Training and Inference}
In the channel charting optimization problem~\eqref{optimization problem}, defining appropriate dissimilarity metrics $d_{ij}$ is crucial. Various metrics have been proposed to quantify distances between channel data, but many fail to account for real-world impairments. As outlined in\cite{stephan2024angle}, these metrics typically fall into two categories: side information-based and channel data-based. In this work, we focus on the latter. Motivated by the importance of channel covariance matrices in MIMO-OFDM systems\cite{li2024downlink} and their robustness against noise and missing channel data, we adopt a spatial covariance matrix-based dissimilarity:
\begin{equation}\label{dissimilarity metric} 
	d_{\text{SCM},ij} = 1 - \frac{\mathrm{trace}\bigl(\mathbf{C}_i^H \mathbf{C}_j\bigr)} {||\mathbf{C}_i||_F||\mathbf{C}_j||_F}, 
\end{equation}
where $\mathbf{C}_i$ and $\mathbf{C}_j$ are the spatial covariance matrices of the $i$-th and $j$-th channel observations, respectively. As illustrated in\cite{stephan2024angle,tenenbaum2000global}, geodesic dissimilarity provides a more rational representation of the distance between channel data and is therefore applied in our work. We adopt a similar procedure as Sec.III.H in\cite{stephan2024angle} to get $d_{\text{G-SCM},ij}$ for our network training.

As illustrated in Fig.~\ref{fig:tdr_model} and according to Sec.~\ref{Section III} A-C, our proposed channel charting function is expressed as:
\begin{align}
\mf z(\cdot) = G_{\boldsymbol{\theta}}(D(F( \cdot))).
\end{align}
Consequently, the training objective of our {\it DTL-based channel charting} is:
\begin{align}
	&\mathcal{L}_{\text{TDL-charting}} = \sum_{i=1}^{N-1}\sum_{j=i+1}^{N} \Big( d_{\text{G-SCM},ij} \notag  \\
	&- \| G_{\boldsymbol{\theta}}(D(F(\boldsymbol{\mathcal{H}}_i))) - 
	G_{\boldsymbol{\theta}}(D(F(\boldsymbol{\mathcal{H}}_j))) \|_2 \Big)^2,
\end{align}
where the network parameters can be learned via Adam optimizer\cite{kingma2014adam}. 

During inference, let $\tilde{\boldsymbol{\mathcal{H}}}$ be an out-of-sample channel data to be charted, and $\boldsymbol{\theta}^{*} = \argmin_{\boldsymbol{\theta}} \mathcal{L}_{\text{TDL-charting}}$. Then, its pseudo-position  is given by:
\begin{equation}
	\tilde{\mathbf{z}} = G_{\boldsymbol{\theta^*}}(D(F(\tilde{\boldsymbol{\mathcal{H}}}))).
\end{equation}

\begin{figure*}[!t]
	\vspace{-2em}
	\centering
	\begin{subfigure}[t]{0.25\linewidth}
		\centering
		\includegraphics[width=\linewidth]{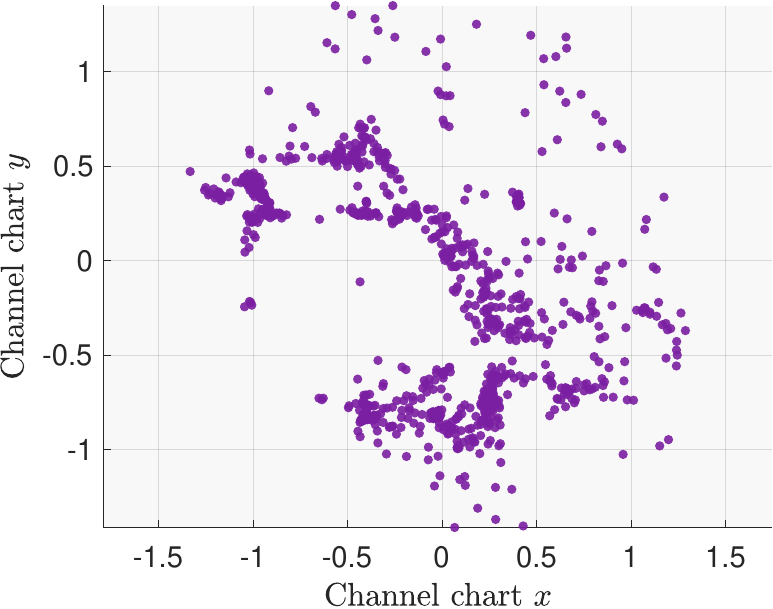}
		\label{fig:sub1}
	\end{subfigure}%
	\hspace{0.04\linewidth}%
	\begin{subfigure}[t]{0.25\linewidth}
		\centering
		\includegraphics[width=\linewidth]{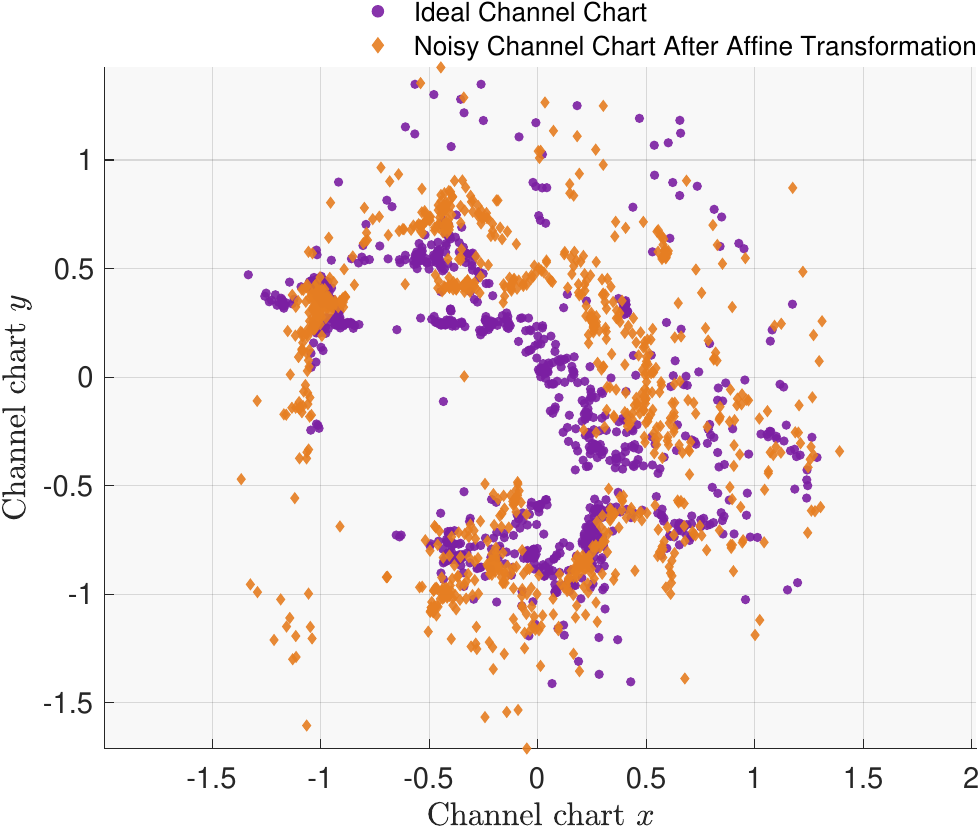}
		\label{fig:sub2}
	\end{subfigure}%
	\hspace{0.04\linewidth}%
	\begin{subfigure}[t]{0.25\linewidth}
		\centering
		\includegraphics[width=\linewidth]{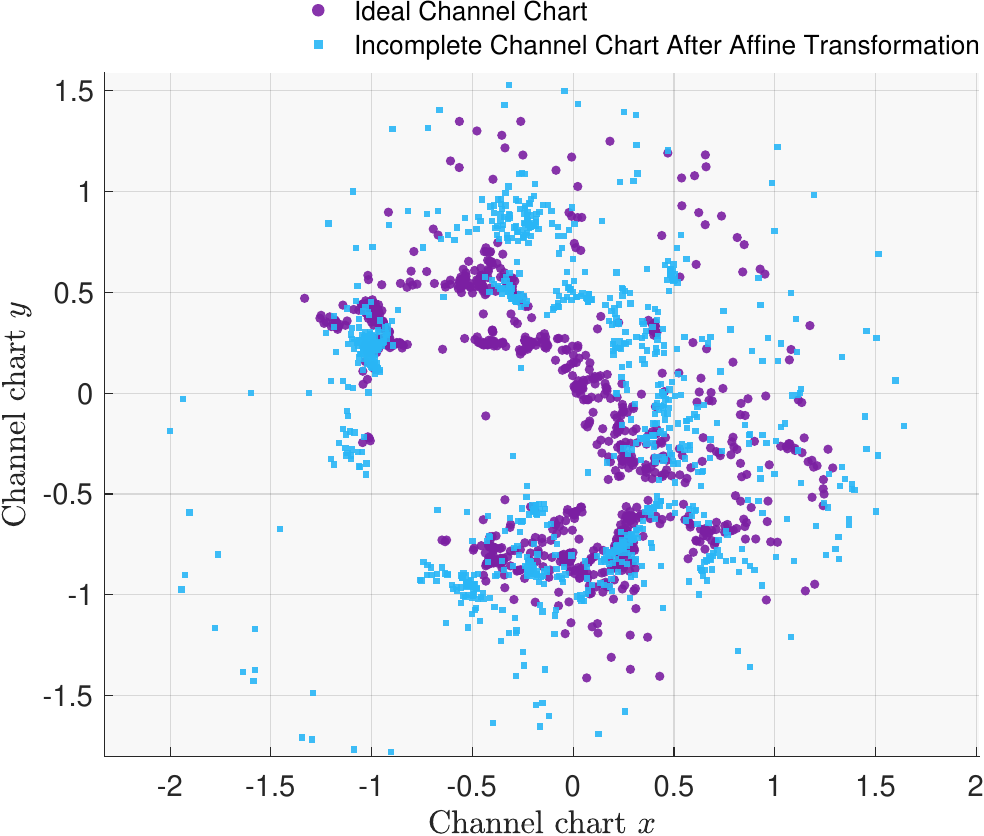}
		\label{fig:sub3}
	\end{subfigure}
	\vspace*{-1em}
	\caption{(Left) Channel chart with ideal data. (Middle) Channel charts using noisy data (SNR = 0 dB, orange points) and ideal data (purple points). (Right) Channel charts via incomplete data (1/17 frequency hopping, blue points) and ideal data (purple points). Despite real-world impairments, both the local and global structures in our DTL-based charting results are well preserved compared to the ideal chart, demonstrating the reliability of our proposed DTL for channel charting.}
	\label{fig:charting results}
	\vspace*{-1em}
\end{figure*}

\section{Experimental Results}
We evaluate our method using Quadriga-simulated channel data consisting 1000 samples. The BS is equipped with a $4\times 8$ dual-polarized rectangular antenna array, yielding $N_{\text{rx}}=32,N_{\text{pol}}=2$. The UE has two transmit antennas ($N_{\text{tx}}=2$), and each channel data includes 408 subcarriers ($N_{\text{sub}}=408$) spaced over a 240kHz bandwidth with a 3.5GHz carrier frequency. For incomplete observations, we consider $1/17$ frequency hopping, i.e., only consecutive $1/17$ subcarrier data can be observed at a time. Channel data is mapped into two-dimensional pseudo-positions, i.e., in~\eqref{formula(3)}, $D=2$.

\textbf{\textit{Baselines}}: We compare 5 channel charting algorithms: two deep learning-based methods—Angle-Delay Profile-based (ADP)\cite{stephan2024angle} and Raw 2\textsuperscript{nd}-moment-based (R2M)\cite{studer2018channel}, and three manifold learning-based methods—UMAP\cite{mcinnes2018umap}, t-SNE\cite{van2008visualizing} and Isomap\cite{tenenbaum2000global}. This paper primarily focuses on deep learning-based methods, as their parametric nature enables the rapid generation of pseudo-positions for new test channels, which is crucial for real-world channel charting applications. In contrast, manifold learning-based methods are non-parametric, making them difficult to extend for out-of-sample charting. These methods are included for comparison to assess the effectiveness of our proposed approach. For deep learning approaches, the batch size, learning rate and training epochs are set to 32, $1\times10^{-3}$ and 300, respectively. 

\textbf{\textit{Performance metrics}}: We adopt three widely used metrics in channel charting: \emph{continuity} (CT), \emph{trustworthiness} (TW), and \emph{Kruskal’s stress} (KS). All three range from 0 to 1. Specifically, CT and TW are optimal at 1, indicating local-neighborhood preservation. KS is optimal at 0 to capture global-structure fidelity. For details of these definitions, see\cite{stephan2024angle}. We also compare the number of trainable parameters for each deep learning method.

\textbf{\textit{Results}}: Tab.~\ref{table1:different channel charting} summarizes the  performance of various charting methods using ideal channel data. Among the deep learning approaches, our proposed DTL achieves the highest CT and TW while yielding the lowest KS, which is attributed to our tensor modeling for feature extraction. Also, our tensor-based neural network leads to significantly fewer parameters, confirming that the tensor computations are effective for network-size reduction. In contrast, UMAP and t-SNE perform the best in terms of CT and TW, but exhibit poor performance in KS. This is because these algorithms prioritize preserving local structure instead of global structure.

 Tab.~\ref{table2: non-ideal channel charting} shows charting results under noisy and incomplete scenarios (SNR=10dB, $1/17$ frequency hopping). Our proposed DTL model still outperforms other deep learning approaches with minimal performance degradation. The large channel features and network parameters in ADP and R2M make them highly susceptible to noise, and they struggle to compute physically meaningful channel features and dissimilarity distances under incomplete observations. In contrast, our DTL model fully leverages low-rank tensor structure in feature design, making it robust and reliable to noisy and incomplete measurements. Additionally, manifold learning algorithms, which only require dissimilarity distance matrix for training, demonstrate stable performance, indicating the inherent robustness of our designed metric~\eqref{dissimilarity metric} in practical settings.
 
 Fig.~\ref{fig:charting results} shows our charting results. Following\cite{stephan2024angle,stahlke2023indoor}, we adopt an optimal affine transformation to the channel chart obtained under noisy and incomplete measurements. Clearly, both local and global structures are preserved well under real-world impairments compared to ideal chart, which indicates the reliability of our proposed DTL  for channel charting.

\begin{table}
	\vspace*{-0.3cm}
	\centering
	\scriptsize 
	\setlength{\tabcolsep}{3pt} 
	\caption{Channel Charting Results For Ideal Channel Data}
	\vspace*{-0.2cm}
	\label{table1:different channel charting}
	\begin{tabular}{c c c c c c c}
		\toprule
		& \textit{Method} & CT$\uparrow$ & TW$\uparrow$ & KS$\downarrow$ & Parameters & Out-of-sample \\
		\midrule
		\multirow{3}{*}{\begin{tabular}{c}Deep \\ learning-based\end{tabular}}
		& \cellcolor{TableGray}\textbf{DTL} & \cellcolor{TableGray}\textbf{0.9922} & \cellcolor{TableGray}\textbf{0.9732} & \cellcolor{TableGray}\textbf{0.1304} & \cellcolor{TableGray}\textbf{11730} & \cellcolor{TableGray}\textbf{Easy} \\
		& ADP & 0.9604 & 0.9264 & 0.2027 & 1.5B  & Easy \\
		& R2M & 0.9152 & 0.9475 & 0.2474 & 985M  & Easy \\
		\midrule
		\multirow{3}{*}{\begin{tabular}{c}Manifold \\ learning-based\end{tabular}}
		& UMAP   & 0.9929 & 0.9866 & 0.3846 & / & Hard \\
		& t-SNE  & 0.9960 & 0.9989 & 0.4180 & / & Hard \\
		& Isomap & 0.9829 & 0.9750 & 0.1375 & / & Hard \\
		\bottomrule
	\end{tabular}
	\vspace*{-0.3cm}
\end{table}

\begin{table}
	\centering
	\scriptsize  
	\setlength{\tabcolsep}{2pt}  
	\caption{Channel Charting Results Under Noisy And Incomplete Observations}
	\vspace*{-0.2cm}
	\label{table2: non-ideal channel charting}
	\begin{tabular}{c c c c c c c c c}
		\toprule
		& \multirow{2}{*}{\textit{Method}} & \multicolumn{3}{c}{SNR: 0dB} & \multicolumn{3}{c}{$1/17$ Frequency Hopping} & \multirow{2}{*}{Out-of-sample} \\ 
		\cline{3-8}\addlinespace[2.5pt]
		&  & CT$\uparrow$ & TW$\uparrow$ & KS$\downarrow$ & CT$\uparrow$ & TW$\uparrow$ & KS$\downarrow$ &  \\
		\midrule
		\multirow{3}{*}{\begin{tabular}{c}Deep \\ learning-based\end{tabular}}
		& \cellcolor{TableGray}\textbf{DTL}  & \cellcolor{TableGray}\textbf{0.9875} & \cellcolor{TableGray}\textbf{0.9727} & \cellcolor{TableGray}\textbf{0.1669} & \cellcolor{TableGray}\textbf{0.9866} & \cellcolor{TableGray}\textbf{0.9665} & \cellcolor{TableGray}\textbf{0.1781} & \cellcolor{TableGray}\textbf{Easy} \\
		& ADP  & 0.8991 & 0.8794 & 0.3595 & / & / & / & Easy \\
		& R2M  & 0.9087 & 0.8797 & 0.5829 & / & / & / & Easy \\
		\midrule
		\multirow{3}{*}{\begin{tabular}{c}Manifold \\ learning-based\end{tabular}}
		& UMAP   & 0.9940 & 0.9857 & 0.3595 & 0.9913 & 0.9848 & 0.4042 & Hard \\
		& t-SNE  & 0.9952 & 0.9977 & 0.4008 & 0.9930 & 0.9967 & 0.3883 & Hard \\
		& Isomap & 0.9776 & 0.9727 & 0.1672 & 0.9757 & 0.9677 & 0.1795 & Hard \\
		\bottomrule
	\end{tabular}
	\vspace*{-0.6cm}
\end{table}

\section{Conclusion}
In this paper, we introduce a deep tensor learning method for reliable channel charting under noisy and incomplete measurements. By exploiting the inherent low-rank structure of MIMO-OFDM channels, we effectively extract features while suppressing noise and handling missing entries. Additionally, our work highlights the importance of integrating domain-specific multi-dimensional data priors into neural networks for processing wireless channel data. Experimental results demonstrate that DTL outperforms state-of-the-art deep learning-based methods with significantly reduced model complexity and is therefore well-suited for practical deployment.
\bibliographystyle{IEEEtran}
\bibliography{references}

\end{document}